\documentclass[onecolumn,showpacs,preprintnumbers]{revtex4}
\usepackage{graphicx}% Include figure files
\usepackage{dcolumn}% Align table columns on decimal point
\usepackage{bm}% bold math

\begin{document}
%\twocolumn[\hsize\textwidth\columnwidth\hsize\csname
%@twocolumnfalse\endcsname
\title{%
\hbox to\hsize{\normalsize\rm
\hfil}
\vskip 36pt Adiabatic evolution of a coupled-qubit Hamiltonian}

\author{V. Corato}
\affiliation{Second University of Naples-INFM, Via Roma 29,I-81031
Aversa, Italy and Istituto di Cibernetica del CNR, via Campi
Flegrei 34, I-80078, Pozzuoli, Italy}

\author{P. Silvestrini}
\affiliation{Second University of Naples-INFM, Via Roma 29,I-81031
Aversa, Italy
and Istituto di Cibernetica del CNR, via Campi Flegrei 34, I-80078,
Pozzuoli, Italy}

\author{L.~Stodolsky}
\affiliation{Max-Planck-Institut f\"ur Physik
(Werner-Heisenberg-Institut),
F\"ohringer Ring 6, 80805 M\"unchen, Germany}

\author{J. Wosiek}
\affiliation{M. Smoluchowksi Institute of Physics, Jagellonian
University, Reymonta 4, 30-059 Cracow, Poland}

\date{\today}% It is always \today, today,
             %  but any date may be explicitly specified

\begin{abstract}

We present a general method for studying coupled qubits driven by
adiabatically changing  external parameters. Extended calculations
are provided for a two-bit Hamiltonian whose eigenstates can be
used as logical states for a quantum CNOT gate. From a numerical
analysis of the stationary Schroedinger equation we find a set of
parameters suitable for representing CNOT, while from a
time-dependent study the conditions for  adiabatic evolution are
determined.
Specializing to a concrete physical system involving SQUIDs, we
determine   reasonable parameters  for experimental purposes. The
dissipation for SQUIDs is discussed by fitting experimental data.
The low
dissipation  obtained  supports the idea that adiabatic operations
could be performed on a time scale shorter than the  decoherence
time.

\end{abstract}

\pacs{03.67.Lx; 03.65.-w; 85.25.Dq; 03.65.Yz}
% PACS, the Physics and Astronomy
% Classification Scheme.
%\keywords{Suggested keywords}%Use showkeys class option if keyword
%display desired
\maketitle

\section{Introduction}

The basic elements for processing quantum information (e.g. to
perform quantum computation \cite{qcomp})
 are quantum bits (qubits), namely two level systems exhibiting
quantum coherence between the states, quantum register
(arrays of qubits) and quantum gates \cite{bar95}. Computations are
performed by the creation of quantum superpositions
 of the qubits and by controlled entanglement of the information on
the qubits \cite{mqc2}. The main goal of any physical
 implementation of a quantum information-processing device is
therefore to control systems of coupled qubit with a phase
coherence time long enough to permit the necessary manipulations.
Various physical systems have been proposed for the
physical implementation of qubits, including photons \cite{photon},
trapped ions \cite{ion},
 spins in nuclear magnetic resonance (NMR) \cite{nmr},
electrodynamics cavities \cite{qed} and semiconductor quantum dots
 \cite{qdot}, to mention some of the most popular.
Other experiments and proposals focus  on
superconducting Josephson devices, where almost macroscopic
(mesoscopic) devices like Josephson junctions \cite{jj}, SQUIDs
\cite{squid} or Cooper pair boxes \cite{cpb},
that is devices fabricated in condensed matter physics, are brought
to behave quantum mechanically. These devices
 promise certain advantages like large-scale integration and
fabrication as well as ease of integration with conventional
 electronics.
 To manipulate qubits quantum  gates
are necessary, that is logic devices capable of operating on linear
combinations of input states.

Among the possible mechanisms for manipulating coupled qubits,
adiabatic procedures \cite{messiah} are of special interest.
Quantum adiabatic evolution provides a natural framework for
solving combinatorial search problems on quantum computers
 \cite{adiabqc}. Any problem which can be recast as the
minimization of an energy function
 (which can then be converted into a quantum Hamiltonian) can
potentially be solved by
 adiabatic quantum computation. General problems have been treated
numerically,
 and studies of a set of Exact Cover instances designed to be hard
have
 shown polynomial behaviour out to instances containing as many as
twenty bits \cite{farhiscie}. Whereas
 a conventional quantum algorithm is implemented as a sequence of
discrete unitary transformations
 that form a quantum circuit involving many energy levels of the
computer, the adiabatic algorithm
 works by keeping the state of the quantum computer close to the
instantaneous ground state of a
 Hamiltonian that varies continuously in time. Therefore, an
imperfect quantum computer implementing
 a conventional quantum algorithm might experience different sorts
of errors than an imperfect adiabatic
 quantum computer. In fact, an adiabatic quantum computer has an
inherent robustness against errors that
 might enhance the usefulness of the adiabatic approach
\cite{childs}. Local operations on single qubits
 (such as NOT) \cite{deco} or two coupled qubits ( such as the
adiabatically controlled CNOT gate we shall discuss)
 \cite{averin}\cite{pla} are also possible, where adiabatic
operations take place
 as a sequence of  discrete  transformations acting
on a few qubits at a time.
 In this line it is important to study a possible trade-off between
the advantages of error reduction
 due to adiabatic evolution and the longer times required for gate
operations.

In this paper we will explain some general principles for studying
adiabatic
SQUID qubit operations focusing particluarly on a  CNOT gate, and
present numerical calculations relevant to
the behavior and design of such systems.

\section{Coupled-qubit Hamiltonian}

In discussing coherence properties of the SQUID under adiabatic
inversion, we have suggested
its interest for the elements of the ``quantum
computer"~\cite{deco}. The single-bit NOT operation can be realized
by
adiabatic
inversion. The next most complicated
 operation is the two-bit operation CNOT, with which a computer
may, in principle, be constructed. CNOT is
 a two-qubit operation and we will try to represent it
by two interacting  double-potential well systems. Qualitatively,
we will use the
procedure of performing an adiabatic  NOT on the first
qubit while trying to influence its behavior by the state of the
second.
We'll find a region
of parameter space where this works.

%\section{ Two Variable Schroedinger Equation}

 In the one dimensional or one SQUID problem  one has a
Schroedinger equation  in the variable $\Phi$  with a kinetic term
and
 the potential term
\begin{equation}
\label{u} U = \frac{{(\Phi  - \Phi ^{ext} )^2 }}{{2L}} - \frac{{I^c
\Phi _0 }}{{2\pi }}\cos (2\pi \Phi /\Phi _0 ).
\end{equation}
which for small $\Phi$ yields a double well potential. $I^c$ and
$L$ are the Josephson critical current and the inductance of the
superconducting ring respectively, while $\Phi^{ext}$ is an applied
external flux. 
When $\Phi^{ext}$ is swept slowly as a function of time, as
explained in ref~\cite{deco}, an adiabatic inversion or level
crossing can be induced, amounting to a realization of the NOT
operation. If the state is originally  in the left potential well
it is transferred to the right well and vice-versa, implying a
reversal of the  current direction in the superconducting ring.

 We now wish to investigate this idea of
quantum gates generated by adiabatic transformations to systems of
more than one variable, in particular for the two variable CNOT
operation. Although we are only
concerned here with two SQUIDS, we
briefly indicate a method valid for many qubits.
The equation for an array of underdamped flux-linked rf SQUIDs is

\begin{equation}\label{phi}
\Phi - \Phi^{ext} = L i
\end{equation}

In this equation $\Phi$ and $i$ are meant as column vectors
representing all the fluxes $\Phi_j$ and currents $i_j$ in the j-th
rf SQUID loop, $\Phi^{ext}$
is the column vector corresponding to the external fluxes, while
$L$ is a matrix
representing the self- and mutual- inductances. The current $i_j$
in j-th ring is expressed in
terms of the capacitance $C_j$ and the
superconducting Josephson current $I_j$ :

\begin{equation}  \label{i}
i_j=-C_j \ddot\Phi_j -\frac{1}{R}\dot \Phi _j- I_j^csin
\Phi_j{2\pi\over\Phi_0}
\end{equation}
where $\Phi_0={hc\over
2e}=2\cdot
10^{-7} G~cm^2$ is the superconducting flux quantum and R is the
effective resistance of the junction.

 An important
property of $L$, by the reciprocity of mutual inductances, is that
$L$ is a symmetric matrix. As we need $i$ for the linear
homogeneous system (LHS) of
Eq[{\ref{i}}], we invert $L$:

\begin{equation}\label{l1}
 i= L^{-1}( \Phi - \Phi^{ext})
\end{equation}

Since $L$ is symmetric, $L^{-1}$ is also symmetric. Neglecting the
dissipative term, we now have Eq~[\ref{i}] as
\begin{equation}  \label{ii}
-\sum_k L^{-1}_{jk}(\Phi -\Phi^{ext})_k - I_j^csin
\Phi_j{2\pi\over\Phi_0}=C_j \ddot\Phi_j
\end{equation}
that can be written as
\begin{equation}  \label{f}
-{\partial U \over \partial \Phi_j}=C_j \ddot\Phi_j
\end{equation}
by introducing the potential

\begin{equation}  \label{U}
U= {1\over 2}\sum_{j,k}L^{-1}_{jk}(\Phi -\Phi^{ext})_k(\Phi
-\Phi^{ext})_j
- ({\Phi_0\over 2\pi}) \Sigma_j I_j^c cos \Phi_j{2\pi\over\Phi_0}
\end{equation}

Finally switching to the reduced dimensionless flux variable
$\phi=\Phi{2\pi\over\Phi_0}$ the last two equations become
\begin{equation}  \label{fa}
-{\partial U \over \partial \phi_j}=({\Phi_0\over 2\pi})^2 C_j
\ddot\phi_j
\end{equation}
\begin{equation}  \label{Ua}
U=({\Phi_0\over 2\pi})^2 {1\over 2}\sum_{j,k}L^{-1}_{jk}(\phi
-\phi^{ext})_k(\phi -\phi^{ext})_j - ({\Phi_0\over 2\pi})\Sigma_j
I_j^c
cos \phi_j
\end{equation}

 %One notes that these results still do not exactly look like
%Eq~[\ref{u}] in that the $cos\phi$ term has the wrong sign.

To make this situation look more symmetric, we follow  recent
practice and introduce the shifts $\phi\rightarrow \phi +\pi$,
$\phi^{ext}\rightarrow \phi^{ext} +\pi$, which move the maximum of
the
$cos\phi$ term to $\phi=0$ \cite{luk95}. This does not affect the
quadratic term, but it must be kept in mind that $\phi^{ext}=0$ now
corresponds to a non-zero applied field. We thus have finally
\begin{equation}  \label{Uaa}
U\rightarrow U=({\Phi_0\over 2\pi})^2 {1\over
2}\sum_{j,k}L^{-1}_{jk}(\phi -\phi^{ext})_k(\phi -\phi^{ext})_j
+({\Phi_0\over 2\pi})\Sigma_j I_j^c cos \phi_j \; ,
\end{equation}
where henceforth we use the shifted variables.

\subsection{Two variables}
We now specialize to two SQUIDS, called 1 and 2, with variables
$\phi_1,\phi_2$. In the absence of mutual inductance, each one can
be thought of as a qubit,
whose dynamics is described
by a double-well potential. For CNOT we
shall think of $\phi_1$ as the target bit and $\phi_2$ as the
control bit. We shall apply a sweeping flux $\phi_1^{ext}$ and
would
like that this sweeping flux induce an adiabatic inversion if the
control bit is $\left| 1 \right\rangle$ (e.g.current flowing
clockwise in SQUID 2), and not induce this inversion if the
control bit is $\left| 0 \right\rangle$(current flowing in the
opposite direction in SQUID 2).

For the two variable system the matrix
$L= \pmatrix{L_1 & L_{12} \cr L_{12} & L_2}$ can be inverted,
giving
\begin{equation}
\label{inv}
L^{-1}= \frac 1{L_1 L_2 - L^2_{12}}
\pmatrix{L_2 & -L_{12} \cr -L_{12} & L_1}
\end{equation}

$L_{12}$, the mutual inductance between the two rings, could be
controlled experimentally by a
 further device with Josephson junctions \cite{carmine} and
  switched off for convenience in performing other operations.

Writing out $U$ we obtain

\begin{equation}  \label{U1}
U= ({\Phi_0\over 2\pi})^2{1\over 2}{1\over
L_1L_2-L^2_{12}}[L_2(\phi_1 -\phi_1^{ext})^2
+L_1(\phi_2 -\phi_2^{ext})^2 -2L_{12}(\phi_1 -\phi_1^{ext})(\phi_2
-\phi_2^{ext})]
+({\Phi_0\over 2\pi})I_1^c cos \phi_1+({\Phi_0\over 2\pi})I_2^c cos
\phi_2
\end{equation}

Let us first look at this potential with the $\phi^{ext}$ zero, and
imagine varying $\phi_1$ at fixed $\phi_2$. For $L_{12}=0$,
we
would simply have the usual symmetric double potential well for
$\phi_1$. Now as $L_{12}$ is turned on, a tilt is
introduced
into this $\phi_1$ potential.  A coupling term $\sim L_{12}\phi_1
 \phi_2$ is added providing a bias in the potential for
$\phi_1$, so that the double well is asymmetric even though
$\phi_1^{ext}=0$. The direction of this bias depends on whether
$\phi_2$ is positive or negative.

If we think of $\phi_2$ as essentially fixed, and now suppose
sweeping $\phi_1^{ext}$, we see that this sweep will either
increase
the asymmetry present in the double well, or decrease it. If the
wells are caused to be further separated, we have no inversion, if
the wells are brought together and cross, we will have an inversion
and
a ``NOT''. Which case occurs  will depend on the sign of $\phi_2$.
The condition for CNOT is accomplished: according to the state of
$\phi_2$ , an inversion takes place or does
not take place in the
$\phi_1$ variable.

%\section{Introduction of dimensionless parameters}

The discussion is more convenient if we introduce dimensionless
parameters
\begin{equation}
\label{defs}
{1\over
L}={\sqrt{L_1L_2} \over
L_1L_2-L^2_{12}}~~~~~~~~\beta_1={2\pi\over \Phi_0} L I_1^c
~~~~~~~~~~\beta_2={2\pi\over \Phi_0} L I_2^c
\end{equation}
as well as the dimensionless inductances
\begin{equation}
\label{defsa}
l_1={L_1\over\sqrt{L_1L_2}}~~~~~~l_2={L_2\over
\sqrt{L_1L_2}}~~~~~~~~~l_{12}={L_{12}\over\sqrt{L_1L_2}}
\end{equation}

We may then write
Eq[\ref{U1}] as
\begin{equation}
\label{U2}
U=({\Phi_0\over 2\pi})^2 {1\over L}\{{1\over
2}[l_1(\phi_1-\phi^{ext}_1)^2 +l_2(\phi_2-\phi^{ext}_2)^2-
2l_{12}(\phi_2-\phi^{ext}_2)(\phi_1-
\phi^{ext}_1)]+\beta_1cos\phi_1+\beta_2cos\phi_2\}
\end{equation}

With this potential the Schroedinger equation for the
wavefunction $\psi(\phi_1,\phi_2,t)$ is
\begin{equation}
\label{sch}
i\dot\psi=\cal{H}\psi
\end{equation}
 with the following Hamiltonian
\begin{equation}
\label{ham}
{\cal H}={-1\over 2C_1 ({\phi_0\over2\pi})^2}{\partial
^2\over\partial \phi^2_1 }+{-1\over 2 C_2({\phi_0\over2\pi})^2}
{\partial ^2\over\partial \phi^2_2 }+U
\end{equation}

Introducing
\begin{equation}\label{c}
C=\sqrt{C_1C_2}
\end{equation}
as the typical capacitance,
 and defining the energy scale
\begin{equation} \label{e01}
 E_0=1/\sqrt{ LC} \;
\end{equation}
$\cal H$ can be cast in the form of an energy times a dimensionless
Hamiltonian
\begin{equation} \label{hh}
 {\cal H}= E_0 H \;.
\end{equation}

\begin{equation}
\label{hama}
H={-1\over 2\mu_1}{\partial ^2\over\partial \phi^2_1 }+{-1\over2
\mu_2}{\partial ^2\over\partial \phi^2_2 }+V
\end{equation}
with $V$
\begin{equation}\label{v}
V=V_0\{{1\over 2}[l_1(\phi_1-\phi^{ext}_1)^2
+l_2(\phi_2-\phi^{ext}_2)^2-
2l_{12}(\phi_2-\phi^{ext}_2)(\phi_1-
\phi^{ext}_1)]+\beta_1cos\phi_1+\beta_2cos\phi_2\}\; ,
\end{equation}
and  the dimensionless parameters

\begin{equation} \label{defsb}
\mu_1= C_1E_0({\Phi_0\over 2\pi})^2
~~~~~~~~~\mu_2=C_2E_0({\Phi_0\over
2\pi})^2~~~~~~~~~V_0=\sqrt{C/L}({\Phi_0\over
2\pi})^2\; .
\end{equation}

\begin{figure}
\includegraphics[width=.8\linewidth]{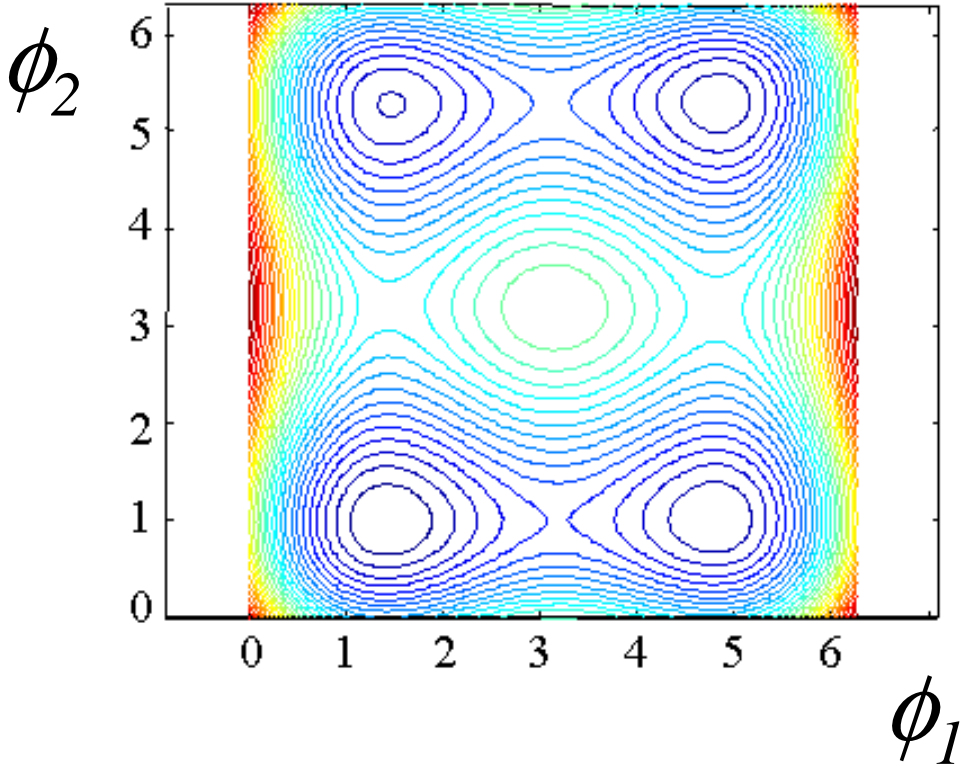}
\caption{\label{fig:pot}(Color online) Potential as in
Eq~[\protect\ref{v}], with its four wells. The coordinate
$\phi_1$(target bit) runs horizontally and $\phi_2$
(control bit) vertically.}
\end{figure}

Fig.~[\ref{fig:pot}] shows the equipotential contours of $V(\phi_1,\phi_2)$, with
its four potential wells.

 We  use natural units: $\hbar=1,c=1, e^2/\hbar c=1/137$, thus
$\Phi_0=hc/2e=\pi \sqrt{137}$, and
 $({\Phi_0 \over 2 \pi})^2=137/4=34.3$; also $\sqrt{farad/henry}=30
$. Using these values
 \begin{equation} \label{defsc}
\mu_1\approx 1030  \sqrt{\bigl({C_1\over
C_2}\bigr)^{1/2}~{C_1/pF\over L/pH}}~~~~~~~~\mu_2\approx 1030
\sqrt{\bigl({C_2\over C_1}\bigr)^{1/2}~{C_2/pF\over
L/pH}}~~~~~~~~~V_0\approx 1030 \sqrt{{C/pF\over L/pH}}  \; .
\end{equation}

Note that these parameters are not
all independent, as the following relation exists
\begin{equation}
\label{non}
\sqrt{\mu_1 \mu_2}= V_0\; .
\end{equation}

That is, the three basic dimensional quantities $C_1$,$C_2$ and $L$
have been exchanged for two dimensionless parameters and the
overall energy scale $E_0$.

\section {CNOT by adiabatic inversion}

With SQUID qubits, the logical states are the flux states of the
superconducting rings ( with one qubit $\left| 0 \right\rangle $ or
$\left| 1 \right\rangle$ with two qubits  $
\left| 0 \right\rangle \left| 1 \right\rangle$, $
\left| 1 \right\rangle \left| 1 \right\rangle$ and so forth),
whereas the Hamiltonian eigenstates are in general a linear
combination of them.
In case of two coupled qubits to form  CNOT, each logical state of
the gate will be
represented by a wave function localized in one of the four
distinct minima of the  potential of Eq[\ref{v}]. The four states
 can be labelled as 1,2,3,4 and placed  in a tableau of the kind
%\begin{equation}\label{tab}
$\pmatrix{
4&3\cr1&2},$
%\end{equation}
where the locations refer to Fig.~[\ref{fig:pot}].
That is, in the tableau  the positions of the
numbers indicate in which well of Fig 1 the state is localized
while the
numbers themselves indicate which  energy eigenstate is meant.
The lowest energy eigenstate is ``1'', and the highest of the four
first states ``4''.

%\subsection {Representation of CNOT}

A CNOT operation is defined by the conditions: A)
the control bit does not change its state, and B) the
target bit is reversed or not reversed, according to whether the
control bit is $\left| 1 \right\rangle$  or  $\left| 0
\right\rangle$.
In the tableau representation,
a physical embodiment of CNOT would be

\begin{equation}\label{map}
\pmatrix{
4&3\cr1&2 }\rightarrow
\pmatrix{4&3\cr2&1}
\end{equation}

 Condition A) on the stability of the control bit is exhibited in
that no states move between the top and bottom row.
Condition B) is realized in that the top row remains unchanged
while the bottom row is ``flipped''.

%\subsection {Adiabatic Operations}
 Realization of operations such as Eq[\ref{map}] can be
accomplished by using adiabatic
processes, thanks to the ``no level
crossing'' behavior of adiabatic evolution. The no-crossing
property  assures that
a state initially in the first, or second, or third,.... energy
level will end up in the first, or second, or third,... energy
level after the adiabatic evolution, while at the same time the
logical state associated
with the level may  be  changing.

One  can proceed as follows: we first search for an initial
Hamiltonian whose variable external parameters
$(\phi_1^{ext},\phi_2^{ext})$
are adjusted to give the first four energy eigenstates localized in
the four different
minima of fig. 1.
 Then, we
search for a final Hamiltonian where another set of
$(\phi_1^{ext},\phi_2^{ext})$,   gives  the tableau on the right of
Eq[\ref{map}].
 In this
 procedure we need only to study  the  {\it stationary}
Schroedinger equation at first. After having determined some
suitable parameter sets,
we shall also study the  full time-dependent
Schroedinger equation
to determine what sweep speed
 is  ``slow'' in order to guarantee adiabatic behavior.

 For the present paper we defer a detailed discussion of the phases
the  states  acquire during the time evolution. These phases
contain dynamic and geometric  contributions \cite{berry} and they
themselves can be used for quantum information processes
\cite{phases}. For a given set of parameters and sweep conditions
the phases can of course be calculated explicitly and in no way
affect the general applicability of the results and procedures
presented here. Calculations of phases  will be reported upon
in future work.

% For the CNOT discussed in this paper,  phases
%don't destroy the universality of the quantum computation, but
%it's necessary to take care of them when one combines the CNOT and
%other elementary
%gates to do a general quantum gate \cite{averin}.

%{\it Switching Behavior:}
We obtain\cite{jacek} the
switching behavior according to  Eq[\ref{map}] by fixing the
control bias $\phi_2^{ext}$ at a relatively high
value, while an adiabatic  sweep of the
target bias
$\phi_1^{ext}$ occurs. This is a generalization of a
NOT~\cite{deco} on $\phi_1$. The presence of the $l_{12}$ coupling
produces an extra bias on
the target bit which "helps or hinders" the NOT operation. The
relatively large bias
 on $\phi_2$ comes from
 condition A): we attempt to ``immobilize'' the control bit despite
the  perturbations communicated
by the sweep of $\phi_1^{ext}$ via $l_{12}$. We therefore
investigate the region $\vert \phi_1^{ext}\vert <<\vert
\phi_2^{ext}\vert $.
If $\phi_2$ is indeed successfully ``immobilized'', it will be
fixed in one of its two potential wells and can have only the
values $\phi_2\approx \pm 1$.
As seen by $\phi_1$, these two states
amount to an extra bias which is added or subtracted to
$\phi_1^{ext}$.
 To linear order (since we take all $\phi_1^{ext},\phi_2^{ext}$
small
compared to $1$ and $\phi_1,\phi_2$ are in the neighborhood of $1$)
and
 introducing the notation $\phi_{1~eff}^{ext}=\phi_1^{ext}\pm
{l_{12}\over
l_1}$ the potential terms
involving    $\phi_1$ in Eq~[\ref{v}] become
\begin{eqnarray} \label{eff}
-2l_1\phi_1\phi_1^{ext}-2l_{12}\phi_1(\pm
1)~~~~~~~~~~~~~~~~~~~~~~~~~~~~~\cr =-
2l_1\phi_1(\phi_1^{ext}\pm {l_{12}\over
l_1})=-2l_1\phi_1\phi_{1~eff}^{ext}\; ,
\end{eqnarray}
so that there is an
effective shift in the  external bias  on $\phi_1$   by $(\pm
{l_{12}\over l_1})$.
According to Eq~[\ref{eff}] the bias condition for switching from
one tableau to another is given
by $\vert \phi_1^{ext}\vert = l_{12}/l_1$.

\begin{figure}
\includegraphics[width=.8\hsize]{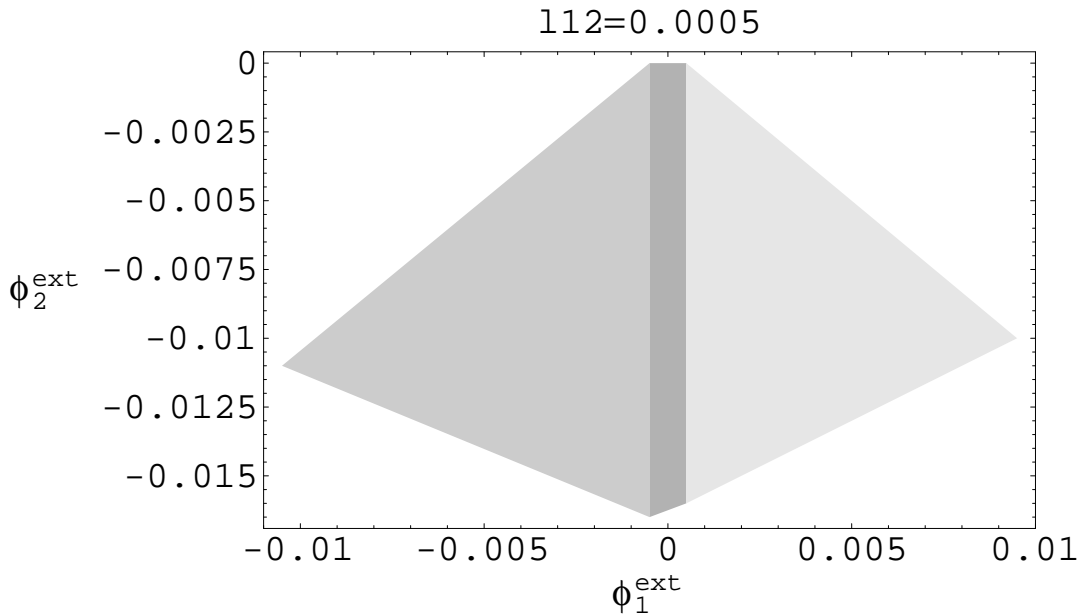}
\caption{\label{fig:2a}A region of the $\phi_1^{ext},\phi_2^{ext}$ plane with a
well
defined set of wavefunctions as explained in the
               text. The coupling  parameter is  $l_{12}=0.0005$,
The other parameters
                 are $l_1=l_2=1,~\beta_1=\beta_2=1.19,~
\mu_1=\mu_2=V_0= 16.3 $.}
\end{figure}

\begin{figure}
\includegraphics[width=.8\hsize]{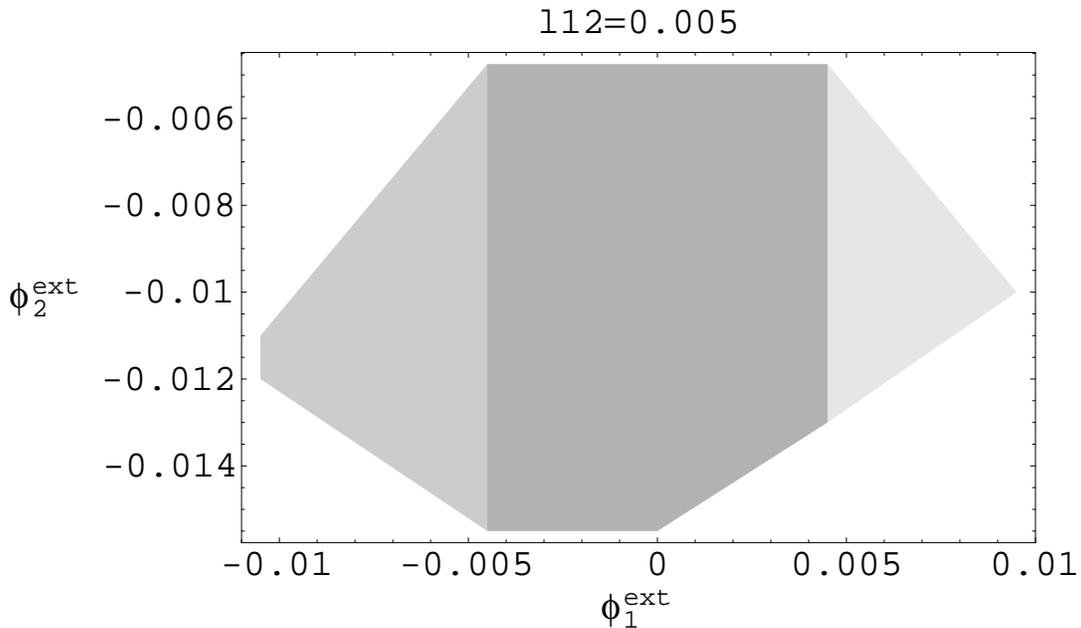}
\caption{\label{fig:2b}As in Fig.2 with   coupling  parameter
$l_{12}=0.005$.}
\end{figure}

%{\it Numerical Results:}
By a numerical study where we  find essentially exact solutions of
the Schroedinger equation for our potential, we find that there is
indeed a region of
$\phi_1^{ext},\phi_2^{ext}$ plane where the eigenstates of
Eq~[\ref{hama}] are well defined and behave as in this description.
These regions are shown in the grey areas of  Figs~[\ref{fig:2a}]
and~[\ref{fig:2b}]. 

By
well defined we mean that the expectation values of
$\phi_1,\phi_2$  are at the location of one of the wells $\approx
(\pm 1,\pm 1)$, that only one of the first four levels is so
localized,  and finally that the dispersion of each coordinate
$\sqrt{<\phi^2> -<\phi>^2}$ is small compared to this expectation
value of $\phi$, that is small compared to 1. For the cases we
 present here the ratio of the dispersion to the expectation
value was in the vicinity of $0.3$.

 We find that the
 three different gray areas of  Fig~[\ref{fig:2a}] and~[\ref{fig:2b}] have well defined
tableaux   as follows:

\begin{equation}
\pmatrix{
3&4\cr1&2 }~~~~~
\pmatrix{4&3\cr1&2}~~~~~
\pmatrix{4&3\cr2&1}
\end{equation}
for the intermediate gray region (left), the darkest region
(center) and the light gray region (right), respectively. On the
order of 10-20  points were used to determine the boundaries in
the figures. These tableaux fit with the description arrived at in
the ``immobilization'' model, where either the top or bottom row
inverts as we go
from the central region  to large $\vert \phi_1^{ext}\vert$. Hence
a
sweep from the central region to the right region will produce the
desired mapping of Eq~[\ref{map}]. Similarly sweeping from the
right
region to the center and from the left region  to the central
region and vice-versa  can also serve as realizations, differing
simply in the assignment of (0,1) for the  bits or the names (1,2)
for the SQUIDs. Note that the switch between tableaux occurs quite
close to
 $\vert \phi_1^{ext}\vert =\vert l_{12}/l_1\vert$,
 as predicted by Eq~[\ref{eff}].

%{\it Adiabatic Condition:}
Concerning the time dependent problem, the  important time
scale in the present context is
$\tau_{adiab}$,  the
shortest time  in
 which an operation can be performed  adiabatically. This time is
relevant with respect
to decoherence and relaxation effects, since the operation must
take
place in  a time short compared to decoherence and relaxation
times.
An  estimate
for $\tau_{adiab}$ for the  NOT operation~\cite{deco} gives
\begin{equation} \label{tab}
\tau _{adiab} =\epsilon  \omega_{tunnel}^{-2}= \epsilon \tau
_{osc}^2\; ,
\end{equation}
where $\epsilon$ is  the asymmetry of the potential
 and $\omega_{tunnel}^{-1}=\tau_{osc}$
 the inverse tunneling energy or oscillation time between the two
states. Since here we also perform a kind of NOT, we expect a
similar relation
to hold, where  $\omega_{tunnel}$ or $\tau
_{osc}^{-1}$ is the smallest level splitting  during the adiabatic
passage and $\epsilon$ may be read off as  the
energy shift of the wells at the beginning and end of our sweep.
 A set of
reasonable SQUID parameters for CNOT are
$L_1=300pH$, $L_2=280pH$, $L_{12}=1.8pH$, $C_1=C_2=0.1pF$ and
$I^{c}_{1}=I^{c}_{2}=1.45\mu A$. Since in frequency units
$E_0\approx
{1\over\sqrt{L/pH ~C/pF}}~ 1000~ GHZ$ these values give $E_0\approx
185~ GHZ$ and $\tau_{adiab}\approx
2.7\cdot
~10^{-9} s$. Preliminary results are in agreement with this
estimate \cite{pla}, and the question  will be studied in more
detail in further numerical work.

\section{Implementation}

\subsection {Dissipation}

As with all discussions of  quantum computation, the important
open questions concern dissipative effects. These include the
decoherence time $\tau_{dec}$, or its inverse the decoherence rate
$D$, as well as relaxation processes , represented by a time
$\tau_{relax}$.

These will affect adiabatic processes as we can see by examining
the  inversion or NOT process for a single qubit.
 With an increasing loss of phase coherence as caused by $D$,
  we expect the situation to become more and more ``classical" and
  finally, when  $D$ is very large, the inversion is
inhibited~\cite{deco}. Furthermore, after the inversion is
completed
there will be some tendency of the upper energy state to relax to
the lower energy state.

\begin{figure}
\includegraphics[width=.5\hsize,angle=-90]{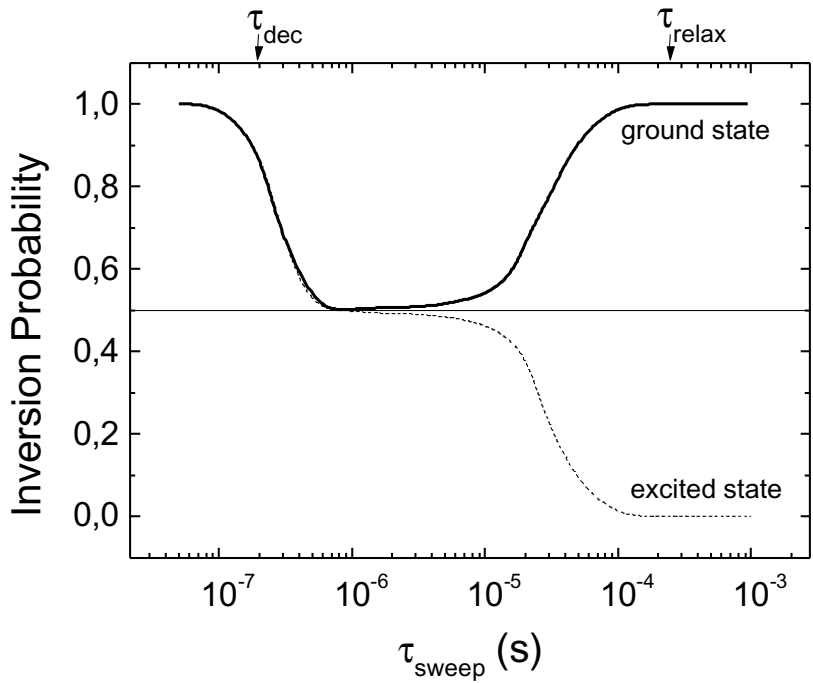}
\caption{\label{fig:P}The probability of inversion as a
function of  sweep time. The inversion probability
is one until the sweep time approaches $\tau_{dec}$, then
it falls to about 0.5. When $\tau_{sweep}$ is increased toward
$\tau_{relax}$ and greater, the system always ends in the  lowest
state, so the inversion probability is one if the
initial state is the ground state, while it falls to zero for
starting in the
excited state.  SQUID parameters used for the simulation were
$\beta_L$=1.2, the loop inductance L=400pH, the junction
capacitance C=0.1pF, the effective resistance R=1M$\Omega$ and the
temperature T=40mK.}
\end{figure}

We illustrate these effects in Fig.~[\ref{fig:P}] where we show the probablity
for  an inversion  as a function of sweep time $\tau_{sweeep}$.
Note that the times involved are much longer than the above
estimate for the adiabatic time
of $2.7\cdot~10^{-9} s$, so the adiabatic condition should be
satisfied.
 Starting at the left, we see that the probability of  inversion
is one
until $\tau_{sweep}$, approaches
$\tau_{dec}$. For $\tau_{sweep}$ much longer than
$\tau_{dec}$ it falls to 0.5, the ``decohered" limit.
 For longer times, as  $\tau_{sweep}$ approaches and passes
$\tau_{relax}$, the final result depends on which state we started
with. Evidently in this limit we will always end up in the ground
state, so the inversion probability is one if we started with the
ground state (solid line) and zero if we started in the excited
state (dotted line).

 The  times used were $\tau_{dec}= 0.2 \mu s$  and
$\tau_{relax}= 0.2 ms$.  The early part of the curve was found
using the plots in Fig 5 of the second of ref\cite{deco} with this
value of $\tau_{dec}$, while the relaxation effects were found
from the calculations shown in Fig 7 of this reference.
 The value
$\tau_{dec}$ used was based on the estimate $D=T/Re^2$
\cite{deco}, with  R=1M$\Omega$ at T=40mK.

It is evident, as exemplified by the latter  formula, that a
necessary and--less evident-- perhaps also a sufficient condition
for quantum behavior of the system  is a low classical dissipation.
In the present context this implies a large value of $R$. Therefore
we present some
experimental data on this point, collected in the thermal regime
for superconducting devices based on SQUIDs \cite{prb}.

In order to evaluate the  dissipation of our system, we
measured the transitions between adjacent flux states of the rf
SQUID as a function of the external flux $\phi^{ext}$. In the
absence of noise, the escape from the metastable well would occur
at a critical value of the external flux $\phi^{c}$. Thermal noise
induces transitions at random values of $\phi^{ext}$ smaller than
$\phi^{c}$, whose probability distribution, namely P($\phi^{ext}$),
was measured by standard ``time fly technique", as explained in
the third reference listed in \cite{jj}.

\begin{figure}
\centering
\includegraphics[width=.6\linewidth,angle=-90]{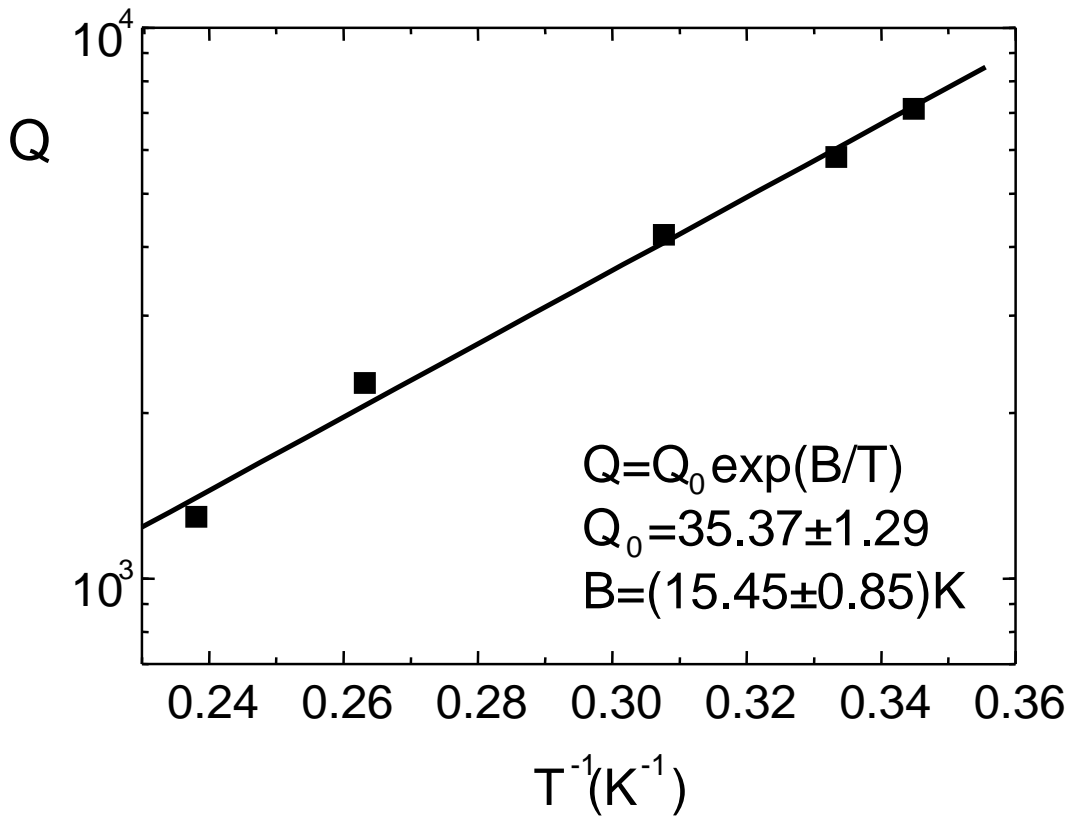}
\caption{\label{fig:Q}Q factor as a function of the inverse of the temperature
obtained by switching flux measurements \cite{prb}. Q increases
exponentially with decreasing temperature, following the
exponential law $Q=Q_{0}e^{B/T}$, with $Q_{0}= 35.37$ and
B=15.45K. This behaviour shows a strongly decreasing
dissipation with  temperature, and  that the dissipation mechanism
is essentially due to the tunneling of thermally activated
quasiparticles in the Josephson junction.}
\end{figure}

By fitting the  data for  P vs $\phi^{ext}$ with Kramers
theory \cite{kramers} in the extremely-low-damping limit, with
L,$I_c$ and C independently measured,
 we can obtain the effective resistance R.
We introduce a dimensionless parameter Q $\sim$ R, defined as
Q=$\omega_0$RC, where $\omega_0$ is the small oscillation
frequency. With decreasing temperature Q  is large,
showing a small dissipation at low temperatures.  The plot of
 Fig.~[\ref{fig:Q}]
indicates an
 exponential increase of R, and using the measured parameters we
obtain R=22k$\Omega$
at T=2.9K. This  is encouraging if the $D=T/Re^2$
estimate is correct, where we needed  1M$\Omega$ at T=40mK.
The
exponential fit  shows that the effective
resistance R is determined by tunneling of thermally activated
quasiparticles, as expected when the external noise has been
filtered out and only the intrinsic dissipation acts.
\subsection{System Design}
On-chip integrated dc-SQUIDs can be used to read out the flux
states
\cite{squid},
while the manipulation of the superconducting qubits can be
controlled by
 on-chip superconducting electronics.  The coupling between the
probe
  and the read-out system could be through a superconducting
transformer. While the intrinsic
  dissipation can be considerably reduced at low temperatures
as explained above \cite{jj}, \cite{squid} a major cause of
difficulty  can be the spurious interaction of the qubit with
readout and control devices.

In a good design, the readout device could be turned off during
the manipulation.
   The technology to control this coupling may need to be
developed, but could be
   reasonably provided by stacked junctions \cite{carmine}, or
small double-junctions
   loop \cite{luk95}, interrupting the coupling transformers. The
coupling is then
    controlled by external current signals. A further possibility
is the development of a fast switch using  simple single flux
quantum circuitry for switching the
 interaction on and off \cite{rsfq}.

%\section {Conclusions}
In conclusion we have presented a general method for studying the
adiabatic evolution of a  Hamiltonian describing a
multi-qubit system, controlled by varying external parameters.
Detailed calculations were provided
 for a two-qubit Hamiltonian, whose eigenstates can be used as
logical states for a quantum CNOT gate.
 From the numerical analysis of the stationary Schroedinger
equation we obtained  sets of parameters
 suitable to perform a CNOT operation, and indicated how a
time-dependent
study determines the limits for  adiabatic evolution.
Specializing to a definite physical system involving SQUIDs, we
identified resaonable values of the parameters, estimated effects
due to dissipation and considered some points of system design.

\section {Acknowledgments}

The authors are grateful to C.Granata and B.Ruggiero for useful
discussions and comments on experimental perspectives. This work
has been partially supported by MIUR-FIRB under project
"Nanocircuiti Superconduttivi".
%%%%%%%%%%%%%%%%%%%%%%%%%%%%%%%%%%%%%%%%%%%%
%%%%%%%%%%%%%%%%%%%%%
%%%%%
%

\end{document}